\documentclass[prd,10pt,onecolumn,showkeys,superscriptaddress,
nofootinbib,
 amsmath,amssymb,aps,
]{revtex4-2}
\usepackage{yfonts}
\usepackage{graphicx}% Include figure files
\usepackage{dcolumn}% Align table columns on decimal point
\usepackage{bm}% bold math
\usepackage{nicefrac}
\usepackage{slashed}
\usepackage{xcolor}
\definecolor{darkblue}{RGB}{0,0,196}
\definecolor{darkgreen}{RGB}{0,120,0}
\definecolor{magenta}{RGB}{255,0,255}
\usepackage[colorlinks=true,linktocpage=true,linkcolor=darkblue,citecolor=red,urlcolor=darkblue]{hyperref}
\usepackage{hyperref}
\usepackage{cancel}
\usepackage{appendix}

\usepackage{etoolbox} % Required for the \ifnumcomp check
\usepackage{yfonts}
\newcommand{\bea}{\begin{eqnarray}}
\newcommand{\eea}{\end{eqnarray}}
\newcommand{\bel}[1]{\begin{eqnarray}\label{#1}}
\newcommand{\eel}{\end{eqnarray}}

\def\LB{\left(}
\def\RB{\right)}
\def\LSB{\left[}
\def\RSB{\right]}

\def\be{\begin{equation}}
\def\ee{\end{equation}}
\def\barr{\begin{array}}
\def\earr{\end{array}}
\def\bfig{\begin{figure}}
\def\efig{\end{figure}}

\newcommand{\EQ}[1]{Eq.~(\ref{#1})}
\newcommand{\EQn}[1]{(\ref{#1})}

\newcommand{\EQSM}[2]{Eqs.~(\ref{#1})--(\ref{#2})}

\newcommand{\CITn}[1]{\citep{#1}}

\def\spin{\,\textgoth{s:}}

\newcommand{\av}{{\boldsymbol a}} 
\newcommand{\bv}{{\boldsymbol b}}

\newcommand{\vv}{{\boldsymbol v}}

\newcommand{\sv}{{\boldsymbol s}}

\newcommand{\pvs}{{\boldsymbol p}}

\newcommand{\dd}[1][]{\mathop{}\!\mathrm{d}^{#1}}

\newcommand{\f}[2]{\frac{#1}{#2}}
\newcommand{\onehalf}{{\nicefrac{1}{2}}}

\newcommand{\tr}{{\rm tr}}  

\def\g5{\gamma_5}

\def\S0iU{{\Sigma}^{0i}}

\def\n0{n_{(0)}}
\def\e0{\varepsilon_{(0)}}
\def\P0{P_{(0)}}

\definecolor{amethyst}{rgb}{0.6, 0.4, 0.8}

\newcommand{\multidot}[1]{%
    \ifnumcomp{#1}{>}{0}{%
        \right.\multidot{\numexpr#1-1\relax}%
    }{}%
}
\newcommand{\multidotleft}[1]{%
    \ifnumcomp{#1}{>}{0}{%
        \left.\multidotleft{\numexpr#1-1\relax}%
    }{}%
}

\newcommand{\beq}{\begin{equation}}
\newcommand{\eeq}{\end{equation}}

\begin{document}

%\preprint{APS/123-QED}

\title{Constructing perfect spin-1 hydrodynamics\\ from Boltzmann to Bose--Einstein statistics}
% Force line breaks with \\

\author{Sudip Kumar Kar}
\email{sudip.kar@doctoral.uj.edu.pl}
\affiliation{Institute of Theoretical Physics, Jagiellonian University, PL-30-348 Krak\'ow, Poland}

\author{Valeriya Mykhaylova}
\email{valeriya.mykhaylova@uj.edu.pl}
\affiliation{Institute of Theoretical Physics, Jagiellonian University, PL-30-348 Krak\'ow, Poland}

\date{\today}
            
\begin{abstract}

We derive thermodynamic currents for a perfect fluid of massive spin-1 particles obeying Bose--Einstein statistics within the Wigner-function approach. Using a covariant spin density matrix, we construct spin-extended equilibrium distributions and obtain the energy–momentum and spin tensors that match, up to second order in polarization, those of spin-1/2 systems and the Boltzmann case. We find that the approach yields a unified description of relativistic spin hydrodynamics independent of statistics and spin representation. Furthermore, the framework fulfills the requirements of the divergence-type theory, with nonlinearly causal and stable dynamical equations. 
\end{abstract}

                              
\maketitle

%%%%%%%%%%%%%%%%%%%%%%%%%%%%%%%%%%%%%%

\section{Introduction}  

In recent years, observables related to spin polarization in heavy-ion collisions have been measured with increasing accuracy~\cite{STAR:2017ckg, Adam:2018ivw, Niida:2018hfw, STAR:2018gyt, STAR:2019erd, STAR:2025jhp, STAR:2025jwc, ALICE:2019aid,  STAR:2022fan, Shen:2024rdq}, indicating the need for a deeper understanding of the mechanisms behind nontrivial spin polarization. To describe the dynamics of spin and, thereby, spin polarization mechanisms in strongly interacting matter, hydrodynamical models must be extended to include spin as an additional degree of freedom. This idea has given rise to spin hydrodynamics for spin-1/2 and spin-1 systems~\CITn{Florkowski:2017ruc, Weickgenannt:2020aaf, Weickgenannt:2021cuo, Bhadury:2020puc, Shi:2020htn, Hattori:2019lfp, Hu:2021pwh, Fang:2025aig, Fukushima:2020ucl, Montenegro:2017rbu}, with a particular focus on the Wigner function formalism~\CITn{Becattini:2013fla, Florkowski:2018ahw, Weickgenannt:2019dks, Hattori:2020gqh, Huang:2020kik, Weickgenannt:2022jes,Wagner:2022gza}. Despite significant progress, many aspects of spin hydrodynamics remain under active construction. For reviews of the main achievements and open challenges see~\CITn{Florkowski:2018fap, Becattini:2020ngo, Mohanty:2021vbt, Becattini:2024uha,Bhadury:2025dzh, Jaiswal:2025ync,Dey:2026epy}. 

For spin-1 systems, tensor polarizabilities and spin alignment of vector mesons have been widely studied experimentally~\cite{STAR:2008lcm,Mohanty:2021vbt,STAR:2022fan,Shen:2024rdq,ALICE:2019aid}. The theoretical investigations of those phenomena evolve within a variety of approaches. These include studies of spin-1 observables based on quark coalescence and fragmentation mechanisms~\CITn{Liang:2004xn, Yang:2017sdk, Sheng:2019kmk, Sheng:2020ghv}, chiral effective theory~\cite{Sheng:2022ffb, Yin:2024dnu, Grossi:2024pyh, Yin:2025gvl}, the Zubarev-response approach~\cite{Li:2025pef}, thermal model~\cite{Banerjee:2026dyp}, quantum kinetic theory with color-field correlators~\cite{Muller:2021hpe}, gradient expansion in relativistic hydrodynamics~\cite{Xia:2020tyd, Li:2022vmb, Kumar:2023ojl}, holography~\cite{Sheng:2024kgg}, and Wigner-function formalism~\cite{Weickgenannt:2022jes,Wagner:2022gza, Wagner:2022eec,Wagner:2023cct,Florkowski:2026ofs,Zhang:2024mhs}.

In our recent work~\cite{Florkowski:2026ofs}, we established a connection between the equilibrium behavior of spin-1/2 and spin-1 particles obeying Boltzmann statistics by exploiting the analogy between the expressions for the spin polarization vector of both systems. %This procedure connects the parameter `$\bm{\omega}$' as defined in a previously proposed form of the equilibrium spin density matrix in~\cite{Xia:2020tyd} to the Lagrange multiplier enforcing spin conservation in the particle rest frame. 
In this study, we extend our methodology to Bose--Einstein statistics. We show that the proposed form of the equilibrium spin density matrix leads to the currents that satisfy general thermodynamic relations, identical to those obtained in the Boltzmann case. To perform a systematic comparison between classical and quantum spin treatments, we additionally evaluate the energy-momentum and spin tensors with a classical description of spin. In both formulations, the resulting thermodynamic expressions have identical structures up to second order in the spin polarization tensor, which coincides with the thermal vorticity in global thermodynamic equilibrium. Following recent works on divergence-type theories for particles with spin-1/2~\cite{Abboud:2025qtg,Bhadury:2025wuh}, we also demonstrate that within our framework, the energy-momentum and spin tensors can be derived from a common generating function. This indicates that our version of perfect spin-1 hydrodynamics is also a divergence-type theory, additionally characterized by fluid-dynamical equations that are nonlinearly causal and stable for both Boltzmann and Bose--Einstein statistics. 

This paper is organized as follows. In Section \ref{sec:SDM}, we review the quantum treatment of a general ensemble of spin states and propose spin-extended equilibrium distribution functions for spin-1 particles obeying Bose--Einstein statistics. In Section~\ref{sec:therm}, we construct thermodynamic currents from the equilibrium distribution function and show that they satisfy the requirements of a divergence-type theory. We subsequently derive macroscopic expressions for the relevant tensors in the limit of small spin polarization and highlight their resemblance to the corresponding results obtained when spin is treated as a classical degree of freedom. Section~\ref{sec:casual} demonstrates that the resulting equilibrium description, besides being a divergence-type theory, is also nonlinearly causal and stable. Finally, we present concluding remarks in Section~\ref{sec:conc}.

\medskip
\textit{Notation and conventions.}~We use natural units, \mbox{$\hbar = c = k_{\rm B} = 1$}, and adopt the mostly minus convention for the metric tensor, $g^{\mu\nu}=\rm{diag}(+1,-1,-1,-1)$. The scalar product of two four-vectors $a$ and $b$ is given by \mbox{$a \cdot b = a^0 b^0 - \av \cdot \bv$}, where bold symbols denote three-vectors. Short-hand notations are introduced for the following types of tensor contractions: $\omega:\omega=\omega_{\mu\nu}\omega^{\mu\nu}$ and $p\cdot\omega\cdot\omega\cdot p=p_\alpha \omega^{\alpha\beta}\omega_{\beta\gamma}p^\gamma$.
 A tilde over an antisymmetric rank-two tensor represents its dual, for example \mbox{${\tilde a}^{\alpha\beta} \equiv (\onehalf) \,\epsilon^{\alpha\beta\gamma\delta} a_{\gamma \delta}$}, where the Levi-Civita symbol $\epsilon^{\alpha\beta\gamma\delta}$ is defined by the convention \mbox{$\epsilon^{0123} =-\epsilon_{0123} = +1$}.   
\bigskip
\section{Spin density matrix}
\label{sec:SDM}
A mixed state of quantum spins is described by a spin density matrix, which is then used to compute the expectation values of various observables. The equilibrium form of this matrix has been debated for many years; see, for example, \cite{Bhadury:2025boe,Becattini:2013fla} for spin-1/2 and \cite{Xia:2020tyd,Florkowski:2026ofs} for spin-1 systems.

We consider a medium of particles with spin 1 and start from the density matrix for the spin states $|1,-1\rangle$, $|1,0\rangle$, and $|1,1\rangle$ defined in the adjoint representation S, in the particle rest frame (PRF)\footnote{The asterisk, whether used as a superscript or subscript, denotes quantities defined in the particle rest frame.} \cite{Florkowski:2026ofs},
\begin{align}
    \rho_{rs*}^S =\frac{1}{3} \left[\delta_{rs} - \frac{3i}{2} {\cal P}^i_* \epsilon_{irs} - \sqrt{6}\, {\cal T}_{rs}^*\right], \label{eq:rhors}
\end{align}
where $r,s ={1,2,3}$ denote spin indices, ${\cal P}^i_*$ are the components of the spin polarization vector, and ${\cal T}_{rs}^*$ are the tensor polarizabilities given, respectively, by~\cite{Leader:2001nas}
\begin{align}
{\cal P}^i_* =& {\rm tr} \left(\rho^S_* S^i \right) \label{eq:Pi},\\[0.5em]
{\cal T}^{ij}_* =& \frac{1}{2}\sqrt{\frac{3}{2}}\left({\rm tr}\left[\rho^S_* (S^i S^j + S^j S^i)\right] -\frac{4}{3}\delta^{ij}\right). \label{eq:Tij}
\end{align}
Note that ${\cal P}^i_*$ and ${\cal T}^{ij}_*$ are, in fact, independent of the representation (inside the trace, the operators $S^i$ can be replaced by any equivalent representation). However, the spin density matrix itself depends on a particular choice of representation and on the corresponding basis of polarization vectors used for the spin states. For a detailed discussion, see~\cite{Florkowski:2026ofs,Palermo:2023cup}. 

In the adjoint representation S, the operators are defined by
\begin{equation}
(S^i)_{jk} = -i \epsilon_{ijk}, 
\label{eq:Sijk}
\end{equation}
 whose spin eigenstates read
\begin{eqnarray}
| e_1 \rangle &=& \frac{1}{\sqrt{2}} (|1,-1 \rangle -   |1,+1 \rangle), \nonumber \\
| e_2 \rangle &=& \frac{i}{\sqrt{2}} (|1,-1 \rangle +   |1,+1 \rangle),  \label{eq:es} \\[0.5em]
| e_3 \rangle &=& |1,0 \rangle . \nonumber
\label{eq:ei}
\end{eqnarray}
Linear combinations of these eigenstates are expressed in terms of the corresponding basis vectors,
\begin{equation}
\epsilon^{i}_{r*} = \delta^i_r = - \delta_{i r},
\label{eq:epsir}
\end{equation}
which are used to construct the general vectors $\lambda_r=l_i \epsilon^i_{r*}$. Multiplication by the basis vectors then yields the spin states $\lambda_r|e_r\rangle$. The index $i$ is treated as a Lorentz index, so that these objects can be interpreted as the polarization vectors of a massive vector field in the PRF, namely $(0,\bm{\epsilon}_{r*})$. In a Lorentz frame moving with velocity $\vv_p$, the general vectors are obtained by boosting the basis vectors from the PRF as follows:
\begin{align}
    \epsilon^{\,\,r}_{\mu}(p)  = L_\mu^{\,\,\nu}(\vv_p)     \epsilon^{\,\,r}_{\nu *}(p).
\end{align}
Here, $L_\mu^{\,\,\nu}(\vv_p)$ is the Lorentz transformation corresponding to the boost $(m,\bm{0})\to (E_{p},\bm{p})$, where \mbox{$E_p = \sqrt{m^2 + \pvs^2}$} denotes the on-shell energy of a particle with mass $m$ and four-momentum $p^\mu=(p^0, \pvs)$. 

With the general polarization vectors at hand, the representation-independent covariant spin density matrix can be defined as
\begin{align}
\rho_{\mu\nu}(x,p) = \epsilon^{\,\,r}_{\mu}(p)
\rho^{S}_{rs*}(x,p)
\epsilon^{\,\,s}_{\nu}(p).
\label{eq:rhomunu}
\end{align}
By introducing Lorentz-boosted forms of the polarization vector and tensor polarizabilities (analogously to the basis vectors $\epsilon^{\,\,r}_{\mu}$),
\begin{align}
{\cal P}^\mu = L^\mu_{\,\,\nu}(\vv_p) {\cal P}^\nu_* , \qquad 
{\cal T}^{\mu\nu} = L^\mu_{\,\,\alpha}(\vv_p)
 L^\nu_{\,\,\beta}(\vv_p) {\cal T}^{\alpha\beta}_*,
\end{align}
we obtain the most general form of the covariant spin-1 density matrix~\cite{Florkowski:2026ofs},
\begin{align}
 \rho_{\mu\nu}(x,p)    =- \frac{1}{3}  \left[\left(g_{\mu\nu}-\frac{p_\mu p_\nu}{m^2}\right) + \frac{3i  \epsilon_{\mu\nu\lambda\rho} {\cal P}^\lambda(x,p)\, p^\rho}{2m}+ \sqrt{6}\,{\cal T}_{\mu\nu}(x,p)\right].
\label{eq:rhomunucov} 
\end{align}

To capture the dynamics of all degrees of freedom, the usual phase-space  distribution of energy states can be modified by the spin density matrix, yielding a new form of the matrix distribution function $\chi_*$, 
\begin{align}
    \rho_{rs*}\to \chi_{rs*}.
\end{align}
For Boltzmann statistics,
\begin{align}
    \chi_{rs*}^{\textrm{B}}(x,p) \sim f_{\rm eq}^{\rm B}(x,p)
\rho^{S}_{rs*}(x,p), 
\label{eq:chi_def}
\end{align} 
where $f_{\rm eq}^{\rm B}(x,p)$ corresponds to spinless equilibrium Boltzmann distribution function, 
\begin{align}
f_{\rm eq}^{\rm B}(x,p) = \exp\left[\xi(x) - p \cdot \beta(x) \right].
\label{eq:f0}
\end{align}
Above, $\xi = \mu/T$ represents the ratio of the chemical potential $\mu$ (associated with the conserved vector meson charges), to the temperature $T$ while $\beta^\mu = u^\mu/T$ with $u^\mu$ being the four-velocity of the fluid. In this study, we set  $\mu=0$ and consider a single particle species with no conserved charges.

Once a spin-dependent distribution function is employed, the expressions for vector polarization ${\cal P}^i_*$ \EQn{eq:Pi} and tensor polarizabilities ${\cal T}^{ij}_*$ \EQn{eq:Tij} are modified by an additional denominator, accounting for extra terms that may affect the normalization,
\begin{gather}
    {\cal P}^i_* = \frac{{\rm tr} \left(\chi_* S^i \right)}{{\rm tr} \left(\chi_*\right)}, \label{eq:Pi2}\\
    {\cal T}^{ij}_* = \frac{1}{2\,{\rm tr} \left(\chi_*\right)}\sqrt{\frac{3}{2}}\left({\rm tr}\left[\chi_* (S^i S^j + S^j S^i)\right] -\frac{4}{3}\delta^{ij}{\rm tr}\left[\chi_* \right]\right). \label{eq:Tij2}
\end{gather}

Following earlier procedure, one may further compute the covariant version of the spin-extended distribution function~\EQn{eq:chi_def}, which can then be connected to the Wigner function via~\cite{Florkowski:2026ofs}
\begin{align}
\mathcal{W}_{\mu\nu}(x,k) &{= -\int \dd P\, \delta^{(4)}(k-p) \chi_{\nu\mu}(x,p) },
\label{eq:gen_WF}
\end{align}
where $k$ is the momentum variable and $\dd P$ is the Lorentz-invariant integration measure,
\begin{align}
\dd P = \frac{{\dd}^3p}{(2\pi)^3 E_p}.
\label{eq:dd P}
\end{align}

In order to describe an ideal gas of spin-1 particles in local equilibrium,
the general form of $ \chi_{rs*}$ given by  Eq.~(\ref{eq:chi_def}) should now be narrowed to capture the thermalization of spin degrees of freedom. Starting from the equilibrium spin density matrix proposed in~\cite{Xia:2020tyd} and adopting Boltzmann statistics, we have devised the equilibrium spin density matrix~\cite{Florkowski:2026ofs}, based on which a matrix distribution in local equilibrium takes the form
\begin{align}
\begin{split}
    {\chi_{rs*}^{\rm B}} &=\exp(-\beta\cdot p +2\av_*\cdot \bm{S})_{rs}\\
    &= g_0^{\rm B}\delta_{rs} + \frac{(\hat{\av}_*\cdot \bm{S})_{rs}}{2} \left(g_+^{\rm B}-g_-^{\rm B}\right)+ \frac{((\hat{\av}_*\cdot \bm{S})^2)_{rs}}{2}\left(g_-^{\rm B}-2g_0^{\rm B}+g_+^{\rm B}\right), \label{eq:integrand}
    \end{split}
\end{align} with
\begin{align}
    g^{\rm B}_-=e^{-p\cdot \beta-2\sqrt{-a^2}},\qquad g_0^{\rm B}= e^{-p\cdot \beta}, \qquad g_+^{\rm B}=e^{-p\cdot \beta+2\sqrt{-a^2}} \label{eq:Bolt}
\end{align}
denoting the modified Boltzmann distribution functions corresponding to the spin states $|1,-1\rangle$, $|1,0\rangle$ and $|1,1\rangle$, respectively. In the above equations, the spin is incorporated through the spin potential $a_\mu$ defined as~\cite{Bhadury:2025boe}
\begin{align}
    a_\mu=-\frac{1}{4m}\epsilon_{\mu\nu\rho\sigma}\omega^{\rho\sigma}p^\nu, \label{eq:amu}
\end{align}
where $\omega^{\rho\sigma}$ is the spin polarization tensor, whose components serve as Lagrange multipliers enforcing spin conservation. In Eq.~(\ref{eq:integrand}), the quantity $\hat{\av}_*\equiv \av_*/|\av_*|$ denotes the spatial direction of the spin potential in the PRF.

We have obtained  the matrix ${\chi_{rs*}^{\rm B}}$ by unifying the kinetic framework for spin-1/2 and spin-1 systems and assuming the interdependence of the polarization vector \EQn{eq:Pi2} and tensor polarizabilities \EQn{eq:Tij2} through the spin potential $a_\mu$. We now extend Eq.~(\ref{eq:integrand}) to Bose--Einstein statistics as follows:
\begin{align}
\begin{split}
    \chi_{rs*}^{\rm BE}&=\left(\exp(\beta\cdot p -2\av_*\cdot \bm{S})-1\right)^{-1}_{rs}\\
    &= g_0^{\rm BE}\delta_{rs} + \frac{(\hat{\av}_*\cdot \bm{S})_{rs}}{2} \left(g_+^{\rm BE}-g_-^{\rm BE}\right)+ \frac{((\hat{\av}_*\cdot \bm{S})^2)_{rs}}{2}\left(g_-^{\rm BE}-2g_0^{\rm BE}+g_+^{\rm BE}\right),  \label{eq:integrandBE}
    \end{split}
\end{align}
with the corresponding distributions incorporating spin degrees of freedom,
\begin{align}
 g_-^{\rm BE}= \frac{1}{e^{\beta\cdot p+2\sqrt{-a^2}}-1},\qquad   g_0^{\rm BE}=\frac{1}{e^{\beta\cdot p}-1}, \qquad g_+^{\rm BE}=\frac{1}{e^{\beta\cdot p-2\sqrt{-a^2}}-1}.
 \label{eq:defs}
\end{align}
The above expressions have been constructed such that they mimic the behavior of spinless Bose--Einstein distribution function, i.e., its reduction to the Boltzmann distribution in the high-temperature and low-density limits.

We further find that, regardless of the applied statistics, the identified form of the spin-extended equilibrium distribution function uniquely determines the polarization vector,
\begin{align}
    \bm{\mathcal{P}}_*= \frac{g_+-g_-}{g_-+ g_0+g_+}\hat{\av}, \label{eq:PolVec2}
\end{align}
and the tensor polarizabilities,
\begin{align}
    \mathcal{T}_*^{ij}= \frac{3{\hat a}^i{\hat a}^j\ - \delta^{ij} }{2 \sqrt{6}  }  \, \frac{g_--2g_0+g_+}{g_- + g_0 + g_+ }, %\label{eq:TensorPol2} 
\end{align}
 in the PRF. Here, the superscripts of the phase-space distributions are omitted, since the expressions hold in both the Boltzmann and Bose--Einstein cases.

\section{Thermodynamics}
\label{sec:therm}
Following the prescription provided in \cite{Weickgenannt:2022jes}, the Wigner function given by \EQ{eq:gen_WF} may be used to compute the relevant thermodynamic quantities and the corresponding conserved currents. It is important to emphasize that one may define the energy-momentum tensor such that its spin and orbital parts are conserved separately (see~\cite{Florkowski:2018fap,Kar:2025qvj} for spin-1/2 particles). Here, we instead adopt the alternative Klein--Gordon version of the tensors, derived from the corresponding Lagrangian for massive vector fields~\cite{Weickgenannt:2022jes},
\begin{align}
    T^{\mu\nu}(x) = \int {\rm d}^4k \,k^\mu k^\nu \, \tr[\mathcal{W}(x,k)],
\label{eq:Tmn}\\
    S^{\lambda,\mu\nu}(x) = 2 i \int {\rm d}^4k\, k^\lambda \, \mathcal{W}^{[\mu\nu]}(x,k), 
\label{eq:Slmn} 
\end{align}
where $\mathcal{W}^{[\mu\nu]} = (\mathcal{W}^{\mu\nu}-\mathcal{W}^{\nu\mu})/2$ is the antisymmetric part of the tensor. With the equilibrium Wigner function~(\ref{eq:gen_WF}), the energy-momentum and the spin tensors can be written, respectively, as
\begin{align}
    T^{\mu\nu}(x) &= \int \dd P \, p^\mu p^\nu \left(g_-+g_0+g_+ \right) \label{eq:Tmunu},\\
     S^{\lambda,\mu\nu}(x) &= \frac{1}{m}\int \dd P\, p^\lambda \frac{g_+-g_-}{\sqrt{-a^2}}  \epsilon^{\mu\nu\alpha\beta} a_\alpha p_\beta,\label{eq:Slambdamunu}
\end{align}
 for both Boltzmann and Bose--Einstein phase-space distributions.

\subsection{Divergence-type structure of the theory}
\label{subsec:3a}
At this point, it is worth mentioning that for spin-1/2 particles, the perfect spin hydrodynamics based on the Boltzmann~\CITn{Bhadury:2025wuh,Abboud:2025qtg} and Bose--Einstein~\CITn{Bhadury:2025wuh} statistics has been found to be a divergence-type theory~\cite{Geroch:1990bw,GEROCH1991394}. To show that this property is also satisfied for particles with spin 1, we define an additional current\footnote{As an example, we perform the calculations using the Bose--Einstein statistics. Similarly, one can consider the Boltzmann case, in which the corresponding current takes the form $\mathcal{N}^\mu_{\rm B} = -\int \dd P \, p^\mu (g_-^{\rm B}+g_0^{\rm B}+g_+^{\rm B})$. }
\begin{align}
    \mathcal{N}^\mu_{\rm BE} = -\int \dd P \, p^\mu \left[\ln (1+g_-^{\rm BE})+\ln(1+g_0^{\rm BE})+\ln(1+g_+^{\rm BE})\right]. \label{eq:Ncal}
\end{align}
By taking the derivative
\begin{align}
    \dd\mathcal{N}^\mu_{\rm BE} &= -\int \dd P \, p^\mu \left(\frac{1}{1+g_-^{\rm BE}}\dd g_-^{\rm BE}+\frac{1}{1+g_0^{\rm BE}}\dd g_0^{\rm BE}+\frac{1}{1+g_+^{\rm BE}}\dd g_+^{\rm BE}\right),
\end{align}
and noticing that
\begin{align}
 \dd g_-^{\rm BE}&=g_-^{\rm BE}(1+g_-^{\rm BE})\left(p^\mu \dd\beta_\mu+\frac{1}{2m\sqrt{-a^2}} \epsilon^{\rho\sigma\beta\kappa}a_{\beta} p_\kappa \dd\omega_{\rho\sigma}\right),\\
    \dd g_0^{\rm BE}\,&=g_0^{\rm BE}\,(1+g_0^{\rm BE})\, p^\mu \dd\beta_\mu,\\
    \dd g_+^{\rm BE}&=g_+^{\rm BE}(1+g_+^{\rm BE})\left(p^\mu \dd\beta_\mu-\frac{1}{2m\sqrt{-a^2}} \epsilon^{\rho\sigma\beta\kappa}a_{\beta} p_\kappa \dd\omega_{\rho\sigma}\right),
\end{align}
we arrive at
\begin{gather}
\begin{split}
    \dd\mathcal{N}^\lambda_{\rm BE} &= -\int \dd P \, p^\lambda \left[g_-^{\rm BE}\left(p^\mu \dd\beta_\mu+\frac{1}{2m\sqrt{-a^2}} \epsilon^{\rho\sigma\beta\kappa}a_{\beta} p_\kappa \dd\omega_{\rho\sigma}\right)\right. \\[0.2em]
    &\hspace{0.4cm}\left.+\, g_0^{\rm BE}\, p^\mu \dd\beta_\mu+g_+^{\rm BE}\left(p^\mu \dd\beta_\mu-\frac{1}{2m\sqrt{-a^2}} \epsilon^{\rho\sigma\beta\kappa}a_{\beta} p_\kappa \dd\omega_{\rho\sigma}\right)\right]\\[0.2em]
    &=-\int \dd P \, p^\lambda p^\mu(g_-^{\rm BE} + g_0^{\rm BE} + g_+^{\rm BE}) \dd \beta_\mu   
    +\frac{1}{2m}\int \dd P \, p^\lambda\frac{g_+^{\rm BE} - g_-^{\rm BE}}{\sqrt{-a^2}} \epsilon^{\rho\sigma\beta\kappa}a_{\beta} p_\kappa \dd \omega_{\rho\sigma}\\[0.2em]
    &=-T^{\lambda\mu}\dd\beta_\mu + \frac12 S^{\lambda,\rho\sigma}\dd\omega_{\rho\sigma}. \label{eq:dNBE}
    \end{split}
\end{gather}
It is clear from the above relation that $T^{\lambda\mu}$ and $S^{\lambda,\rho\sigma}$ defined in \EQSM{eq:Tmn}{eq:Slmn} can be obtained from a common generating function $\mathcal{N}^\lambda$, varied with respect to the corresponding field. This shows that our version of perfect spin-1 hydrodynamics described by Bose--Einstein statistics is a divergence-type theory. We have also verified that this property is satisfied for spin-1 system described by Boltzmann statistics.

\subsection{Small polarization expansion}
Given the small polarization observed in heavy-ion collisions, the components of the spin polarization tensor $\omega^{\mu\nu}$ are expected to be correspondingly small. Additionally, $\omega^{\mu\nu}$ is a dimensionless quantity in natural units, which makes it a suitable expansion parameter.

First, we consider the case of classical statistics, which leads to the energy-momentum tensor of the form~\cite{Florkowski:2026ofs}
\begin{align}
    T^{\mu\nu}_{\rm B}(x) &= \int \dd P \, p^\mu p^\nu f_{\rm eq}^{\rm B} \left[2\cosh\left(2\sqrt{-a^2}\right)+1 \right], \label{eq:TmunuB}
\end{align}
with $f_{\rm eq}^{\rm B}$ defined in \EQ{eq:f0}.
Using~\EQ{eq:amu}, the magnitude of the spin potential $a^\mu$ can be evaluated explicitly as
\begin{align}
    a^2=-\frac{\omega:\omega}{8} - \frac{p\cdot\omega\cdot\omega\cdot p}{4m^2}. \label{eq:a2}
\end{align}
As has been shown in~\cite{Drogosz:2024gzv}, it is convenient to parameterize the spin potential in terms of vectors $k_\mu$ and $\omega_\mu$, which are orthogonal to the flow,
\begin{align}
    \omega_{\mu\nu}=k_\mu u_{\nu}-k_{\nu}u_{\mu} +\epsilon_{\mu\nu\alpha\beta}u^\alpha \omega^\beta. \label{eq:omega_param}
\end{align}
For the subsequent calculations, it is also useful to define
\begin{align}
    t^\mu &= t^{\mu\nu} k_\nu = \epsilon^{\mu \nu \alpha \beta} k_\nu u_\alpha \omega_\beta. 
\end{align}
The parametrization \EQn{eq:omega_param} yields $\omega:\omega=2(k^2-\omega^2)$. Using this expression, together with~\EQ{eq:a2}, we expand the hyperbolic function in~\EQ{eq:TmunuB} up to second order in the spin potential,
\begin{align}
\begin{split}
    T^{\mu\nu}_{\rm B}(x)&=3\int \dd P \, p^\mu p^\nu g_0^{\rm B} +(k^2-\omega^2)\int \dd P \, p^\mu p^\nu g_0^{\rm B} +\frac{\omega_{\alpha\beta}\omega^\beta_{\,\,\gamma}}{m^2}\int \dd P \,p^\alpha p^\gamma p^\mu p^\nu g_0^{\rm B}+\mathcal{O}(\omega^4). \label{eq:Tmunu39}
    \end{split}
\end{align}
We now define the quantities
\begin{align}
    Z^{\mu\nu\cdots}&=\int \dd P p^\mu p^\nu\cdots g_0, \label{eq:Zmunu}
\end{align}
where the ellipsis denotes higher-order moments of the distribution function. In particular, the explicit form of the second moment reads
\begin{align}
    Z^{\mu\nu} &= \int \dd P \, p^\mu p^\nu g_0 = \frac{\epsilon_0}{3}u^\mu u^\nu-\frac{P_0}{3}\Delta^{\mu\nu}, \label{eq:Zmunu2}
\end{align}
where $\epsilon_0$ and  $P_0$ are, respectively, the energy density and pressure of spinless particles, while \mbox{$\Delta^{\mu\nu}=g^{\mu\nu}-u^\mu u^\nu$} is the operator projecting on the space orthogonal to the flow. We note that \EQSM{eq:Zmunu}{eq:Zmunu2} are valid for both types of the  statistics considered.

Using the definition \EQn{eq:Zmunu}, we rewrite \EQ{eq:Tmunu39} --  where the last term then takes the form $Z^{\alpha\gamma\mu\nu}\omega_{\alpha\beta}\omega^\beta_{\,\,\gamma}$ --  and explicitly perform the tensor contractions to obtain the energy-momentum tensor
\begin{align}
    T^{\mu\nu}_{\rm B}(x) &= (\epsilon_{0}+{\epsilon}_{2}^{k}+{\epsilon}_{2}^{\omega})  u^\mu u^\nu - ({P}_{0}+{P}^{k}_{2}+{P}^{\omega}_{2}) \Delta^{\mu\nu} + P_{k} k^\mu k^\nu  + P_{\omega}\omega^\mu \omega^\nu  + P_{t} (t^\mu u^\nu + t^\nu u^\nu), \label{eq:Thybrid}
\end{align}
with scalar coefficients
\begin{align}
    {\epsilon}_{0} &=\frac{3z^{2}T^{4}}{2\pi^{2}} \LSB z K_{3}(z)-K_{2}(z) \RSB, \label{eq:coefeps}\\ 
{\epsilon}_{2}^{k} &= -\frac{z T^{4}}{ \pi^{2}}\LSB z K_{2}(z) + 5K_{3}(z) \RSB k^{2}, \\
{\epsilon}_{2}^{\omega} &= -\f{2zT^{4}}{\pi^{2}} \LSB z K_{2}(z)+(z^{2}+10)K_{3}(z) \RSB\omega^{2},
\end{align}
and
\begin{align}
    {P_{0}}&=\frac{3z^{2}T^{4}}{2\pi^{2}} K_{2}(z),\\
{P}^{k}_{2}&=-\frac{2zT^{4}}{ \pi^{2}}K_{3}(z)k^{2}, \quad {P}^{\omega}_{2} = -\frac{zT^{4}}{2\pi^{2}} \LSB z K_{2}(z)+4K_{3}(z) \RSB\omega^{2},\\
P_{t}&=\f{zT^{4}}{\pi^{2}} \LSB K_{3}(z)-zK_{4}(z) \RSB, \quad  P_{k}=P_{\omega}=-\f{z \, T^{4}}{\pi^{2}}  K_{3}(z), \label{eq:coefP}
\end{align}
where $K_n(z)$ for $n=\{2,3,4\}$ and $z=m/T$ are the modified Bessel functions of the second kind.

The spin tensor involving the Boltzmann distribution is expressed as~\cite{Florkowski:2026ofs}
\begin{align}
    S^{\lambda,\mu\nu}_{\rm B}(x) = 2\int \dd P\, p^\lambda f_{\rm eq}\frac{\sinh \left(2\sqrt{-a^2}\right)}{m\sqrt{-a^2}}  \epsilon^{\mu\nu\alpha\beta} a_\alpha p_\beta,
\end{align}
and its second-order expansion yields only terms linear in the spin potential,
\begin{align}
    S^{\lambda,\mu\nu}_{\rm B}(x) &=2\omega^{\mu\nu}Z^\lambda+ \frac{2}{m^2}\omega^{\nu}_{\,\,\alpha}Z^{\lambda\alpha\mu} -\frac{2}{m^2}\omega^{\mu}_{\,\,\alpha}Z^{\lambda\alpha\nu}. \label{Slmn2}
\end{align}
After the explicit tensor contractions, \EQ{Slmn2} takes the form
\begin{align}
    S^{\lambda,\mu\nu}_{\rm B}(x) &=  u^\lambda\LSB A \LB k^\mu u^\nu - k^\nu u^\mu \RB + A_1 t^{\mu\nu}\RSB  + \frac{A}{2}\LB t^{\lambda \mu} u^\nu - t^{\lambda \nu} u^\mu +\Delta^{\lambda \mu} k^\nu - \Delta^{\lambda \nu} k^\mu\RB, \label{eq:Shybrid}
\end{align}
with
\begin{align}
    A &= -\f{2z T^{3}}{\pi^{2}}K_{3}(z), \label{eq:A}\\
     A_{1} &= \f{zT^{3}}{\pi^{2}} \LSB zK_{2}(z)+2K_{3}(z)\RSB. \label{eq:A1}
\end{align}

The energy-momentum and spin tensors obtained in Eqs.~(\ref{eq:Thybrid}) and (\ref{eq:Shybrid}) have the structures identical to those obtained for particles with spin 1/2 described by Boltzmann statistics~\cite{Drogosz:2024gzv}. However, the corresponding scalars \mbox{\EQn{eq:coefeps}-\EQn{eq:coefP}} and \EQn{eq:A}-\EQn{eq:A1} for spin-1 and spin-1/2 differ by constant prefactors related to different numbers of the possible spin states.

In the case of Bose--Einstein statistics, we perform the expansion around the spinless distribution function $g_0^{\rm BE}$,
\begin{align}
    g_+^{\rm BE}=g_0^{\rm BE}+2g_0^{\rm BE}(1+g_0^{\rm BE})\sqrt{-a^2}-2g_0^{\rm BE}(1+g_0^{\rm BE})(1+2g_0^{\rm BE})a^2+\mathcal{O}(a^4),\\
    g_-^{\rm BE}=g_0^{\rm BE}-2g_0^{\rm BE}(1+g_0^{\rm BE})\sqrt{-a^2}-2g_0^{\rm BE}(1+g_0^{\rm BE})(1+2g_0^{\rm BE})a^2+\mathcal{O}(a^4).
\end{align}
As a result, the energy-momentum tensor given by \EQ{eq:Tmunu} becomes
\begin{align}
    T^{\mu\nu}_{\rm BE}(x) &= \int \dd P \, p^\mu p^\nu \left[3g_0^{\rm BE}+g_0^{\rm BE}(1+g_0^{\rm BE})(1+2g_0^{\rm BE})\left(\frac{\omega:\omega}{2} + \frac{p\cdot\omega\cdot\omega\cdot p}{m^2}\right)\right] \\
    &=3Z^{\mu\nu} + \frac{\omega:\omega}{2}Z_2^{\mu\nu}+ \frac{\omega_{\alpha\beta}\omega^\beta_{\,\,\gamma}}{m^2}Z_2^{\alpha\gamma\mu\nu}, \label{eq:TmnBE}
\end{align}
where $Z^{\mu\nu}$ is given by \EQ{eq:Zmunu}, while $Z_2^{\mu\nu}$ and $Z_2^{\alpha\gamma\mu\nu}$ are given by
\begin{align}
    Z_2^{\mu\nu\cdots}= \int \dd P \, p^\mu p^\nu\cdots g_0^{\rm BE}(1+g_0^{\rm BE})(1+2g_0^{\rm BE}).
\end{align}

Expanding the spin tensor \EQn{eq:Slambdamunu} in terms of $g_0^{\rm BE}$, we get 
\begin{align}
    S^{\lambda,\mu\nu}_{\rm BE}(x) &= \frac{2}{m^2}\int \dd P\, p^\lambda 4g_0(1+g_0)  (m^2\omega^{\mu\nu}+p^\mu\omega^{\nu}_{\,\,\alpha} p^\alpha-p^\nu\omega^{\mu}_{\,\,\alpha} p^\alpha) \\
    &=2\omega^{\mu\nu}Z^\lambda_2 +\frac{2}{m^2}\omega^{\nu}_{\,\,\alpha} Z^{\lambda\mu\alpha}_2-\frac{2}{m^2}\omega^{\mu}_{\,\,\alpha} Z^{\lambda\nu\alpha}_2. \label{eq:SlmnBE}
\end{align}
Equations \EQn{eq:TmnBE} and \EQn{eq:SlmnBE} can be straightforwardly compared with the corresponding expressions obtained in the earlier study of the spin-1/2 system by expanding the spinless Fermi--Dirac distribution (see Eqs. (60) and (63) in~\cite{Drogosz:2024gzv}). We observe identical tensor structures, up to different constant prefactors associated with the different spin representations of spin-1/2 fermions and spin-1 bosons.

We additionally note that when the expansion is carried out to second order, the energy-momentum tensor contains even powers of the spin polarization, whereas the spin tensor is linear in $\omega$. This is consistent with generalized thermodynamic relations, in which the spin tensor is multiplied by the spin polarization tensor, leading to terms quadratic in $\omega$ (for example, see \EQ{eq:dNBE}).

\subsection{Classical spin description for spin-1 particles}
The treatment of spin as a classical degree of freedom~\cite{Mathisson:1937zz,Mathisson:2010opl} simplifies certain aspects of hydrodynamics calculations~\cite{Florkowski:2018fap}. In this approach, the distribution function $f(x,p,s)$ is extended to spin space by introducing the internal angular momentum tensor of the particle, $s^{\alpha\beta}$, which accounts for spin degrees of freedom via
\begin{align}
    s^{\alpha\beta} = \frac{1}{m} \epsilon^{\alpha\beta\gamma\delta} p_\gamma s_\delta. \label{eq:sab}
\end{align}
Here, $s^\delta$ denotes the particle spin four-vector, which has only spatial components in the PRF, $s^\alpha_*=(0,\sv_*)$. This implies that the particle spin is orthogonal to its four-momentum in any reference frame,
\begin{align}
s\cdot p=0, \label{eq:sdotp}
\end{align}
which allows us to invert \EQ{eq:sab} and obtain 
\begin{align}
    s^{\alpha} = \frac{1}{2m} \, \epsilon^{\alpha\beta\gamma\delta} p_\beta s_{\gamma \delta}.
\end{align}
The magnitude of $ s^{\alpha}$ equals to the eigenvalue of the quadratic Casimir operator in the spin-1 irreducible representation,
\begin{align}
    s^\alpha s_\alpha = \spin^2 = 1(1+1) = 2. \label{eq:ssquared}
\end{align}

In order to evaluate the phase-space integrals over the spin configurations satisfying the conditions \EQn{eq:sdotp} and \EQn{eq:ssquared}, we define the integration measure 
\begin{align}
    \dd S=\frac{3}{2}\frac{m}{\pi\spin } {\rm d}^4 s\,\delta(s\cdot s +\spin^2)\delta(p\cdot s),
\end{align}
such that it additionally satisfies $\int \dd S=3$ for three possible orientations of spin 1. 

With this setup, we define the energy-momentum and spin tensors as 
\bel{eq:Tmunu_cl}
T^{\mu \nu}_{\rm eq} = \int \dd P \,\dd S \, p^\mu p^\nu \, f_{\rm eq}(x,p,s)
\eel 
and 
\bel{eq:Slmunu_cl}
\hspace{-0.5cm}S^{\lambda, \mu\nu}_{\rm eq}\!&=&\!\!\int \!\dd P \, \dd S \, \, p^\lambda \, s^{\mu \nu} f_{\rm eq}(x,p,s),
\eel
respectively. Both expressions hold for Boltzmann and Bose--Einstein statistics, in which the corresponding spin-extended equilibrium distribution functions are given by
\begin{align}
f_{\rm eq}^{\rm B}(x,p,s)&=\exp \LB -p \cdot \beta(x)  + \frac{1}{2} \, \omega(x) : s \RB,\\
    f_{\rm eq}^{\rm BE}(x,p,s) &= \LSB
\exp \LB p \cdot \beta(x)  -  \frac{1}{2} \, \omega(x) : s \RB -1 \RSB^{-1}.
\end{align}
It can be straightforwardly shown that the framework with a classical description of spin successfully reproduces the results obtained from the quantum spin treatment up to second order in $\omega$.

\section{Causality and stability}
\label{sec:casual}

The compatibility of our framework with a divergence-type theory (see Sec.~\ref{subsec:3a}) greatly simplifies the analysis of well-posedness, stability and causality of the resulting fluid-dynamical equations. To test causality and stability, we consider the four-vector~\cite{Abboud:2025qtg,Bhadury:2025wuh}
\begin{align}
M^{\lambda}\equiv M^{\lambda AB} \cal{Z}_A Z_B,
    \label{eq:Mlambda}
\end{align}
where $\cal{Z}_A=(\cal{Z}, \cal{Z}_\mu, \cal{Z}_{\mu\nu})$ are non-vanishing real multi-index objects that cycle through tensors of different ranks. It has already been shown for spin-1/2 systems~\cite{Abboud:2025qtg,Bhadury:2025wuh} that the future-directed timelike nature of $M^\lambda$ ensures that the evolution equations admit nonlinearly causal and stable solutions~\cite{Geroch:1990bw, Gavassino:2022roi,GEROCH1991394}. The four-vector $M^{\lambda}$ can be also derived from
\begin{align}
    M^\lambda = \hat{\cal{Z}}^2 {\cal N}^\lambda, \label{eq:MlambdaGEN}
\end{align}
where we introduce the linear differential operator
\begin{align}
 {\hat{\cal{Z}}} =  {\cal{Z}}\frac{\partial}{\partial\xi} - {\cal{Z}}_{\mu}\frac{\partial}{\partial\beta_{\mu}} + \frac{1}{2} {\cal{Z}}_{\mu\nu} \frac{\partial}{\partial \omega_{\mu\nu}}. \label{eq:Zhat}
\end{align}

\subsection{Boltzmann statistics}
We now proceed to the explicit evaluation of \EQ{eq:MlambdaGEN} for Boltzmann statistics. As noted earlier, 
\begin{align}
    \mathcal{N}^\mu_{\rm B} = -\int \dd P \, p^\mu (g_-^{\rm B}+g_0^{\rm B}+g_+^{\rm B}),
\end{align}
with spin-extended distribution functions provided by \EQ{eq:Bolt}.

The first application of the operator $\hat{{\cal{Z}}}$ gives
\begin{align}
  {\hat{\cal{Z}}} {\cal N}^\lambda_{\rm B}=\int \dd P \, p^\mu \left[ \left({ {\cal{Z}} \cdot p}+\frac{a\cdot\tilde{\cal{Z}}\cdot p}{m\sqrt{-a^2}}\right)g_+^{\rm B}+({\cal{Z}} \cdot p)g_0^{\rm B}+\left({\cal{Z}}\cdot p-\frac{a\cdot\tilde{{\cal{Z}}}\cdot p}{m\sqrt{-a^2}}\right)g_-^{\rm B}\right],
\end{align}
while a second application yields
\begin{align}
\begin{split}
   {\hat {{\cal{Z}}}}^2  {\cal N}^\lambda_{\rm B} &=M^\lambda_{\rm B}=  \int \dd P \, p^\mu \left[\left({\cal{Z}} \cdot p+\frac{a\cdot\tilde{{\cal{Z}} }\cdot p}{m\sqrt{-a^2}}\right)^2g_+^{\rm B}+ \left({\cal{Z}} \cdot p-\frac{a\cdot\tilde{{\cal{Z}} }\cdot p}{m\sqrt{-a^2}}\right)^2g_-^{\rm B}\right.\\
   &\hspace{0.3cm}\left.+({\cal{Z}}\cdot p)^2g_0^{\rm B}+\frac{1}{m\sqrt{-a^2}}\left(-(\tilde{{\cal{Z}} }\cdot p)^2 + \frac{(a\cdot\tilde{{\cal{Z}} }\cdot p)^2}{a^2}\right)(g_+^{\rm B}-g_-^{\rm B})\right].
   \end{split}
\end{align}
In the above equation, the integrand is positive, indicating that $M^\lambda_{\rm B}$ is future-oriented and timelike. Thus, the hydrodynamic equations based on Boltzmann statistics are nonlinearly causal and stable.

\subsection{Bose--Einstein statistics}

We repeat the procedure described in the previous section for the Bose-Einstein case. Acting with the operator ${\hat{{\cal{Z}}}}$ \EQn{eq:Zhat} on the four-vector $ {\cal N}^\lambda_{\rm BE}$ defined in \EQ{eq:Ncal}, we obtain
\begin{align}
\begin{split}
 {\hat {\mathcal{Z}}}  {\cal N}^\lambda_{\rm BE}= &\left( {\mathcal{Z}}\frac{\partial}{\partial\xi} - {\mathcal{Z}}_{\mu}\frac{\partial}{\partial\beta_{\mu}} + \frac{1}{2} {\mathcal{Z}}_{\mu\nu} \frac{\partial}{\partial \omega_{\mu\nu}}\right){\cal N}^\lambda_{\rm BE}\\
    =& \int \dd P \, p^\mu \left[\left({\mathcal{Z}}\cdot p+\frac{a\cdot\tilde{{\mathcal{Z}}}\cdot p}{m\sqrt{-a^2}}\right)g_+^{\rm BE}+({\mathcal{Z}}\cdot p)g_0^{\rm BE}+\left({\mathcal{Z}}\cdot p-\frac{a\cdot\tilde{{\mathcal{Z}}}\cdot p}{m\sqrt{-a^2}}\right)g_-^{\rm BE}\right].
    \end{split}
\end{align}
A second application leads to
\begin{align}
\begin{split}
  {\hat {\cal{Z}}}^2  {\cal N}^\lambda_{\rm BE}=   & \int \dd P \, p^\mu \left[\left({\cal{Z}}\cdot p+\frac{a\cdot\tilde{{\cal{Z}}}\cdot p}{m\sqrt{-a^2}}\right)^2g_+^{\rm BE}(1+g_+^{\rm BE})+ \left({\cal{Z}}\cdot p-\frac{a\cdot\tilde{{\cal{Z}}}\cdot p}{m\sqrt{-a^2}}\right)^2g_-^{\rm BE}(1+g_-^{\rm BE})\right.\\
    &\left.+\,({\cal{Z}}\cdot p)^2g_0^{\rm BE}+\frac{1}{m\sqrt{-a^2}}\left(-(\tilde{{\cal{Z}}}\cdot p)^2 + \frac{(a\cdot\tilde{{\cal{Z}}}\cdot p)^2}{a^2}\right)(g_+^{\rm BE}-g_-^{\rm BE})\right]. \label{eq:Z2NBe}
    \end{split}
\end{align}
We note that the combinations $g_+^{\rm BE}(1+g_+^{\rm BE})$, $g_-^{\rm BE}(1+g_-^{\rm BE})$ and $(g_+^{\rm BE}-g_-^{\rm BE})$ are all positive. In addition, the quantity
\begin{align}
    -(\tilde{{\cal{Z}}}\cdot p)^2 + \frac{(a\cdot\tilde{{\cal{Z}}}\cdot p)^2}{a^2}
\end{align}
takes values between $0$ and $1$. Therefore, the integrand in \EQ{eq:Z2NBe} is positive, ensuring that $M^\lambda$ is future-directed and timelike, and hence that the theory is causal and stable.

\section{Conclusions}
\label{sec:conc}
In this work, we extended our previously developed framework for the equilibrium description of spin-1 fluids from Boltzmann to Bose--Einstein statistics. 
The methodology is based on the mechanism through which the spin-extended Bose--Einstein distribution reduces to the corresponding Boltzmann form in the dilute limit (exactly as for spinless systems). This construction also resembles our earlier study of equilibriated spin-1/2 fluids obeying Boltzmann or Fermi-Dirac statistics. We found that, irrespective of the chosen statistics and spin value, the macroscopic currents have identical structure and satisfy the same generalized thermodynamic relations.

To facilitate comparison between quantum and classical treatments of spin, we also formulated a classical-spin description for spin-1 particles. Expanding the relevant currents up to second order in the spin polarization tensor, we found that the classical and quantum approaches lead to identical tensor structures. This result further supports the interpretation of spin as an effective classical degree of freedom in the regime of small polarization, relevant for heavy-ion collisions.

Using the corresponding equilibrium Wigner function, we derived the energy-momentum and spin tensors and demonstrated that they can be obtained from a common generating function. Consequently, the resulting perfect spin-1 hydrodynamics has the structure of a divergence-type theory for both Boltzmann and Bose--Einstein statistics. This extends our analogous results obtained previously for spin-1/2 systems, highlighting the universality of the underlying thermodynamic framework.

Finally, exploiting the divergence-type structure of the theory, we showed that the  resulting hydrodynamic equations satisfy the properties of non-linear causality and stability. Therefore, our results for the equilibrium thermodynamics of spin-1 fluids appear to be largely independent of the underlying particle statistics while retaining the mathematical properties necessary for a consistent hydrodynamic formulation.
\medskip
\begin{acknowledgments}
This work was supported in part by the National Science Centre, Poland (NCN) Grant No.~2022/47/B/ST2/01372. We thank Wojciech Florkowski for useful comments.
\end{acknowledgments}

\bibliography{references}

\end{document}